\documentclass{aa}  
\usepackage{graphicx}
\usepackage{txfonts}
\usepackage{natbib}
\usepackage{color}
\usepackage{tabularx}
\usepackage{tabu}
\pdfoutput=1 

\begin{document} 

\title{Revealing the structure of the lensed quasar Q 0957+561}
\subtitle{III. SMBH mass via gravitational redshift}

\author{C. Fian\inst{\ref{inst1}}, E. Mediavilla\inst{\ref{inst3},{\ref{inst4}}}, J. Jim\'enez-Vicente\inst{\ref{inst5},{\ref{inst6}}}, V. Motta\inst{\ref{inst7}}, J. A. Mu\~noz\inst{\ref{inst8},\ref{inst9}}, D. Chelouche\inst{\ref{inst2}}, A. Hanslmeier\inst{\ref{inst10}}}
\institute{School of Physics and Astronomy and Wise Observatory, Raymond and Beverly Sackler Faculty of Exact Sciences, Tel-Aviv University, Tel-Aviv, Israel\label{inst1} \and Instituto de Astrof\'{\i}sica de Canarias, V\'{\i}a L\'actea S/N, La Laguna 38200, Tenerife, Spain\label{inst3} \and Departamento de Astrof\'{\i}sica, Universidad de la Laguna, La Laguna 38200, Tenerife, Spain\label{inst4} \and Departamento de F\'{\i}sica Te\'orica y del Cosmos, Universidad de Granada, Campus de Fuentenueva, 18071 Granada, Spain\label{inst5} \and Instituto Carlos I de F\'{\i}sica Te\'orica y Computacional, Universidad de Granada, 18071 Granada, Spain\label{inst6} \and Instituto de F\'{\i}sica y Astronom\'{\i}a, Universidad de Valpara\'{\i}so, Avda. Gran Breta\~na 1111, Playa Ancha, Valpara\'{\i}so 2360102, Chile\label{inst7} \and Departamento de Astronom\'{i}a y Astrof\'{i}sica, Universidad de Valencia, E-46100 Burjassot, Valencia, Spain\label{inst8} \and Observatorio Astron\'{o}mico, Universidad de Valencia, E-46980 Paterna, Valencia, Spain\label{inst9} \and Haifa Research Center for Theoretical Physics and Astrophysics, University of Haifa, Haifa, Israel\label{inst2} \and Institute of Physics (IGAM), University of Graz, Universit{\"a}tsplatz 5, 8010, Graz, Austria\label{inst10}}


\abstract
{}
{We intend to use the impact of microlensing on the Fe III $\lambda\lambda$2039-2113 emission line blend along with a measure of its gravitational redshift to estimate the mass of the quasar's central supermassive black hole (SMBH).}
{We fit the Fe III feature in multiple spectroscopic observations between 2008 and 2016 of the gravitationally lensed quasar Q 0957+561 with relatively high signal-to-noise ratios (at the adequate wavelength). Based on  the statistics of microlensing magnifications, we used a Bayesian method to derive the size of its emitting region.}
{The Fe III $\lambda\lambda$2039-2113 spectral feature appears systematically redshifted in all epochs of observation by a value of $\Delta \lambda \sim$17\AA\ on average. We find clear differences in the shape of the Fe III line blend between images A and B. Measuring the strength of those magnitude differences, we conclude that this blend may arise from a region of half-light radius of $R_{1/2}\sim15$ lt-days, which is in good agreement with the accretion disk dimensions for this system. We obtain a mass for the central SMBH of $M_{BH} = 1.5_{-0.5}^{+0.5}\times 10^9 M_\odot$, consistent within uncertainties with previous mass estimates based on the virial theorem. The relatively small uncertainties in the mass determination ($<35\%$) make this method a compelling alternative to other existing techniques (e.g., the virial plus reverberation mapping based size) for measuring black hole masses. Combining the Fe III $\lambda\lambda$2039-2113 redshift based method with the virial, we estimate a virial factor in the $f \sim 1.2-1.7$ range for this system.}
{}

\titlerunning{SMBH mass of Q 0957+561 via gravitational redshift}
\authorrunning{Fian et al.} 
\maketitle

\section{Introduction \label{intro}}
The experimental determination of the masses of the supermassive black holes (SMBH) that reside at the centers of active galactic nuclei (AGN) can be related, via the virial theorem, to the velocity of the gas moving under the gravitational potential of the SMBH, $\Delta V$. An estimate of the distance of the gas to the SMBH, $R$, is needed in addition. Finally, the largely unknown geometry and dynamics of the gas distribution should also be taken into account through a coefficient, $f$, the virial factor, which greatly affects to the mass determination (see, e.g., \citealt{Peterson1999,Peterson2000,Fromerth2000,Krolik2001,McLure2001}):
\begin{equation}
M_{BH} = f {(\Delta V)^2R\over G}.  
\end{equation}

The velocity estimate,  $\Delta V$, can be inferred from the Doppler broadening of the emission lines associated to the ionized gas. The specific definition of $\Delta V$ is not straightforward, thus, several options that, in turn, affect the value of $f$ are usually considered (see, e.g., \citealt{Collin2006}). Furthermore, the complexity of the broad emission lines, which often present several components and substructure, makes difficult the experimental determination of the line widths. The size of the emitting region can be obtained from reverberation mapping (RM, see reviews by \citealt{Peterson1993,Peterson2006}) or from the size luminosity, R-L, scaling relationship (calibrated using RM; see, e.g., \citealt{Wu2004,Dalla2020,Fonseca2020,Yu2020}) or from microlensing (\citealt{Wambsganss2006}). As a consequence of the experimental difficulties and the scarcity of knowledge regarding $f$, virial masses have a typical uncertainty of about 0.4 dex.\\

An alternative way to determine SMBH masses is the accretion disk fitting technique. Based on an adequate selection of the model parameters (specifically of the SMBH spin and disk inclination), this method has the capacity to obtain results that are compatible with virial-based ones, with similar uncertainties (\citealt{Campitiello2019}).\\

The masses of SMBH can also be measured with a new method based on the redshift of an UV iron feature, namely, the Fe III$\lambda\lambda$2039-2113 blend. This method has been successfully applied (i.e., providing mass estimates that are statistically compatible with virial masses) to a sample of ten lensed quasars (\citealt{Mediavilla2018}) and to another sample of ten non-lensed quasars (\citealt{Mediavilla2019}) observed with X-shooter at the Very Large Telescope (\citealt{Capellupo2015,Capellupo2016}). Although the redshift method is basically insensitive to the geometry and the presence of nongravitational forces, it is also tied to a distance estimator:
\begin{equation}
\label{mass}
M_{BH}={2 c^2\over3 G}{\left(\Delta \lambda\over \lambda\right)_{FeIII}}{R_{FeIII}},
\end{equation}

 \noindent where $\Delta \lambda/\lambda$ is the redshift of the Fe III$\lambda\lambda$2039-2113 emission lines and $R_{FeIII}$ the size of the region emitting this blend. In a number of cases, \citet{Mediavilla2018,Mediavilla2019} used an average size for $R_{FeIII}$ inferred from microlensing (\citealt{Fian2018}), scaled to each object using the $R\propto \lambda L_\lambda^\alpha$ relationship with $\alpha\sim 0.5$ (\citealt{Wandel1999}). This approach has led to good results but, certainly, the application of the method to objects with individual measurements of the size is important for validating the technique. This was done in \citet{Mediavilla2018} using UV spectra of NGC 5548 to estimate the RM size of the Fe III$\lambda\lambda$2039-2113 blend. Here, we use a collection of spectra from a well-known gravitationally lensed quasar, Q 0957+561, to estimate (based on microlensing variability) the size of the region emitting the Fe III$\lambda\lambda$2039-2113 feature in this object.\\ 

The paper is organized as follows. In \S \ref{2} and \S \ref{3} we analyze the spectroscopic data, obtaining the redshift for the Fe III$\lambda\lambda$2039-2113 blend and the microlensing-based size to determine the SMBH mass. In \S \ref{4} we discuss these results and compare them with other measurements. Finally, our main conclusions are presented in \S \ref{5}.

\section{Data analysis and redshift measurements}\label{2}
\begin{figure}
\centering
\includegraphics[width=9.1cm]{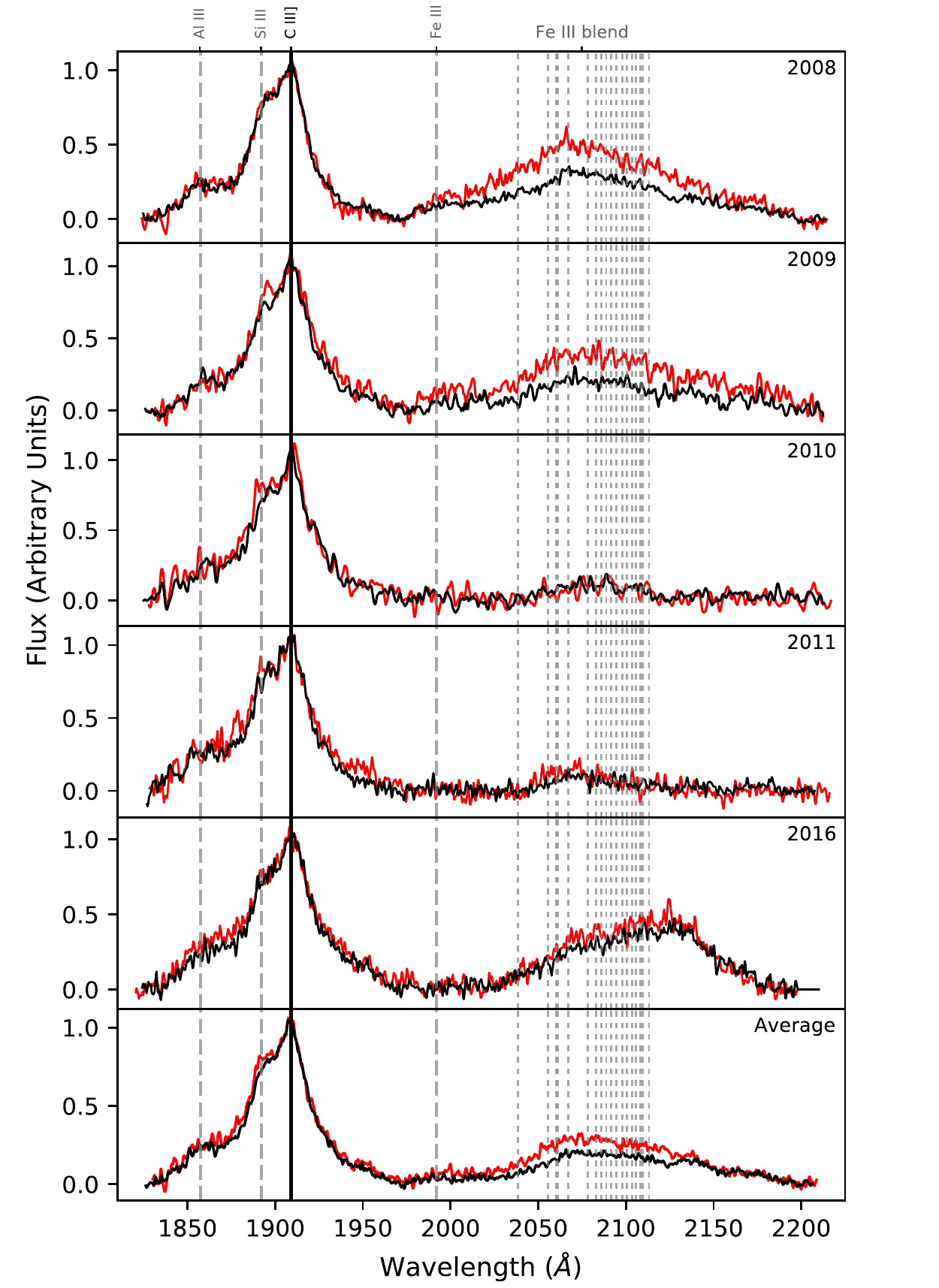}
\caption{Superimposed spectra showing the C III] emission line and the FeIII $\lambda\lambda$2039-2113 feature for image A (black) and image B (red) in different epochs. We subtracted the continuum and matched the C III] line cores of both images. The ordinate is in arbitrary units of flux.}
\label{epochs}
\end{figure}

The spectroscopic data analyzed in this work have different origins (see Table \ref{data}). We model the Fe III $\lambda\lambda$2039-2113 emission feature in the spectrum of the lensed quasar at five different epochs of observations between 2008 and 2016. First, we fit a straight line ($y=a\lambda+b$) to the continuum, one on either side of the C III] emission line and another one defined in two windows at the blue and red side of the Fe III blend. When convenient, the windows used to fit the continuum have been adapted to the shape of the continuum, which can change from epoch to epoch. Finally, we subtract the continuum from the spectrum and normalize the continuum-subtracted spectra to match the core of the C III] emission line defined by the flux within a narrow interval ($\pm$ 4\AA) centered on the peak of the line. This normalization removes the effects of the macro-magnification produced by the lens galaxy and the differential extinction between the images. Thus, the cores of the emission lines are used as a reference that is little affected by microlensing and intrinsic variability (see \citealt{Guerras2013,Fian2018}) as they arise from a significantly larger region than the wings. The continuum-subtracted and core matched spectra in the wavelength regions around the C III] emission line and the (microlensed and redshifted) Fe III $\lambda\lambda$2039-2113 emission feature can be seen in Figure \ref{epochs}.\\

\begin{table}[h]
\renewcommand{\arraystretch}{1.15}
\tabcolsep=0cm
\caption{Spectroscopic data.}
\begin{tabu}to 0.49\textwidth {X[l]X[l]X[l]}

\hline
\hline
Date & Facility & Reference \\
\hline 
01/2008 & MMT & \citealt{Motta2012} \\
01/2009 & NOT & GLENDAMA$^{*}$\\
03/2010 & NOT & GLENDAMA$^{*}$\\
12/2011 & NOT & GLENDAMA$^{*}$\\
03/2016 & WHT & \citealt{Fian2021} \\
\hline 
\end{tabu}
\label{data}
*see \citealt{GilMerino2018}
\end{table}

\begin{figure}
\centering
\includegraphics[width=9cm]{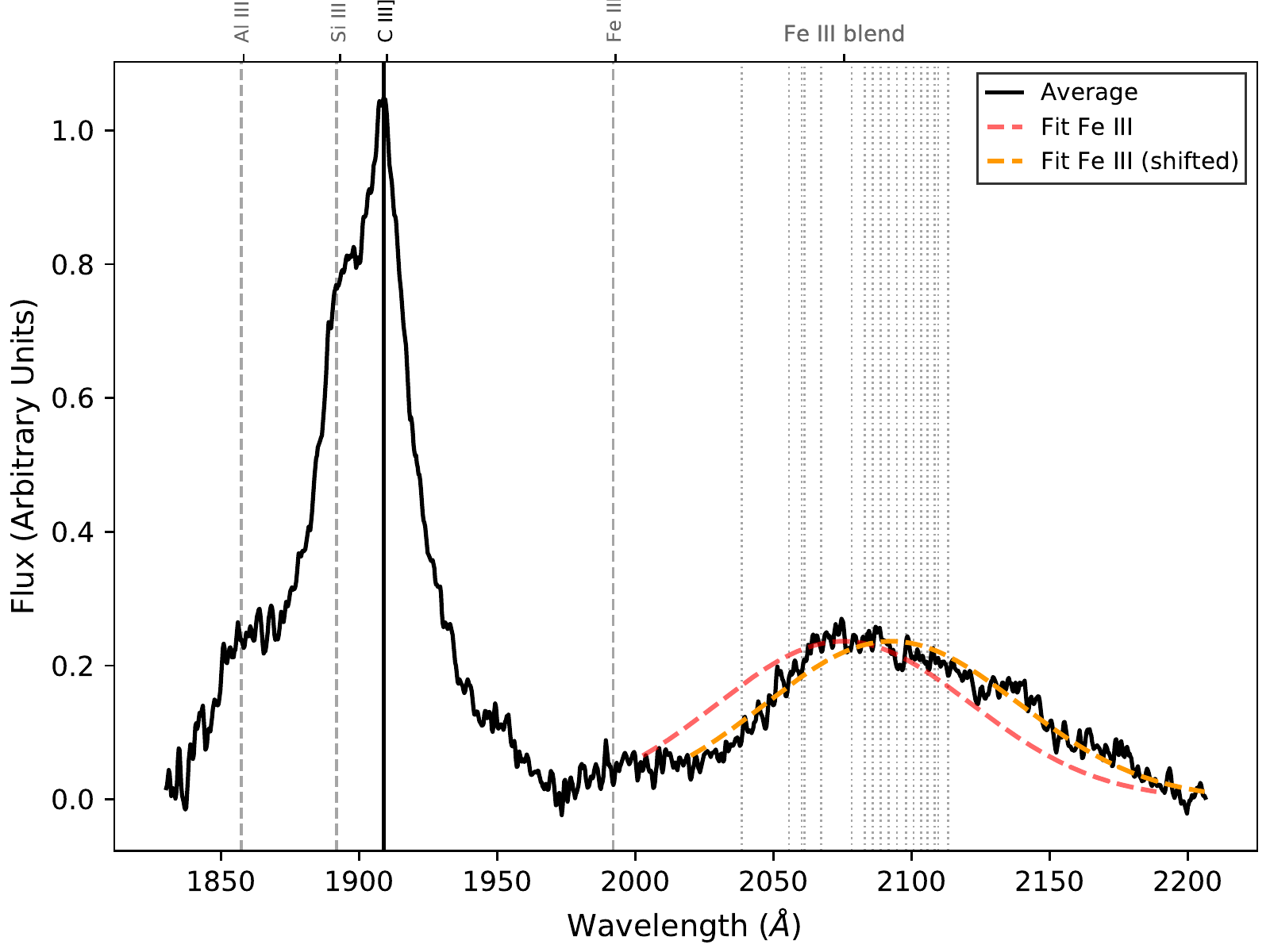}
\caption{Average spectrum (black) and fit to the Fe III $\lambda\lambda$2039-2113 blend (dashed red curve). Vertical dotted lines are located at the wavelengths corresponding to the Fe III lines of the \citet{Vestergaard2001} template. The fit has been shifted by $\sim 17$\AA\ to match the template rest frame (dashed orange curve). The ordinate is in arbitrary units of flux.}
\label{average}
\end{figure}

We fit the Fe III$\lambda\lambda$2039-2113 blend using a template of 19 single Fe III lines of fixed relative amplitudes as provided by \citet{Vestergaard2001}. The (Gaussian) lines are broadened with the same width, $\sigma$ (=FWHM/2.35), globally  shifted in wavelength by $\Delta \lambda$, and their relative amplitudes multiplied by a global scale factor, $K_{scale}$. These are the three free parameters used to fit the blend in both images in different epochs. The Fe III $\lambda\lambda$2039-2113 feature is relatively free of contamination from lines of other species and from Figure \ref{average} we can see that the template of \citet{Vestergaard2001} is able to reproduce well the shape of it when a redshift is applied. The average ($\pm$ standard error in the mean) of this systematic redshift is $\langle \Delta \lambda \rangle = 17.4\pm4.0 \AA$, in agreement with the findings of \citet{Mediavilla2018,Mediavilla2019}, who also included one epoch of this system in their analysis.

\section{Estimation of the Bayesian source size and SMBH mass}\label{3}

We analyzed three different wavelength regions to measure the microlensing-induced changes in the UV Fe III emission line blend. In the first case, we estimate the magnitude difference, $\Delta m$, in each epoch over a fixed wavelength range of $\lambda\lambda$2039-2113. We use the following statistics to calculate the magnitude difference at each wavelength $x$ between the images A and B:
\begin{equation}
\Delta m_x = w_x * (B_x-A_x),
\end{equation}

 \noindent with weights $w_x = \sqrt{\langle B_x+A_x\rangle /(B_x+A_x)}$. These weights are selected to equalize the typical deviations of the differences, that is, to take into account the fact that the calculation of the difference between two emission lines may imply the use of data with very different S/N ratios (after removing the continuum, the data close to the peak have high S/N ratios as compared with the data in the extreme wings). From the mean value in a given wavelength interval, $\langle \Delta m_x \rangle$, we compute the magnitude difference between the images, $\Delta m = \langle \Delta m_x\rangle$. In the second case, we shift the wavelength window according to the estimated redshift in that epoch, and in the third case, we apply a mean redshift of $\langle \Delta \lambda \rangle = 17.4 \AA$ to the wavelength interval of case one. The estimated magnitude differences are similar in all three cases and are given for each epoch in Table \ref{magdiff}.
\begin{table}[h]
\tabcolsep=0.cm
\renewcommand{\arraystretch}{1.2}
\caption{Magnitude differences $\Delta m$ for a fixed range ($\lambda\lambda$2039-2113), individually shifted range and a globally shifted ($\sim$+17\AA) range ($\lambda\lambda$2056-2130).}
\begin{tabu}to 0.49\textwidth {X[c]X[c]X[c]X[c]}

\hline 
\hline
Epoch & Fixed & Shifted & +17.4$\AA$\\
\hline 
I & $-0.39\pm0.03$ & $-0.39\pm0.03$ & $-0.39\pm0.04$ \\
II & $-0.53\pm0.08$ & $-0.54\pm0.08$ & $-0.53\pm0.09$ \\
III & $-0.01\pm0.23$ & $-0.00\pm0.25$ & $-0.03\pm0.27$ \\
IV & $-0.21\pm0.28$ & $-0.15\pm0.31$ & $-0.04\pm0.29$ \\
V & $-0.15\pm0.06$ & $-0.12\pm0.03$ & $-0.15\pm0.04$ \\
\hline 
\end{tabu}
\label{magdiff}
\end{table}

We use these estimates in combination with simulations of the microlensing effect in Q 0957+561 to infer the more likely size of the region emitting the Fe III blend. Our simulations are based in 2000$\times$2000 pixel microlensing magnification maps, generated at the positions of the images using the inverse polygon mapping (IPM) method described in \citet{Mediavilla2006,Mediavilla2011}. The general characteristics of the magnification maps are determined (for each quasar image) by the local convergence, $\kappa$, and the local shear, $\gamma$. To compute the magnification maps, we used $\kappa_A=0.20$, $\kappa_B = 1.03$, $\gamma_A=0.15$, and $\gamma_B=0.91$, obtained by fitting a singular isothermal sphere with an external shear (SIS+$\gamma_e$), such as might be generated by the tide from a neighboring galaxy or cluster, which reproduce the flux ratio and the coordinates of the images (\citealt{Mediavilla2009}). The local convergence is proportional to the surface mass density and can be divided into $\kappa = \kappa_c+\kappa_\star$, where $\kappa_c$ is the convergence due to continuously distributed matter (i.e., dark matter) and $\kappa_\star$ is due to the stellar-mass point lenses (i.e., stars in the galaxy). The produced maps span 400$\times$400 lt-days$^2$ on the source plane, with a pixel size of 0.2 lt-days. We assume a mean stellar mass of $M=0.3M_\odot$ and for the fraction of mass in stars we use $\alpha=10\%$. All linear sizes can be rescaled with the square root of the microlens mass $\sqrt{M/M_\odot}$. To simulate the effect of finite sources, we model the luminosity profile of the emitting region as a Gaussian ($I \propto exp(-R^2/2r_s^2)$) and the magnifications experienced by a source of size $r_s$ are then found by convolving the magnification maps with the Gaussian profiles of sigma $r_s$. We used a linear grid for the source sizes, spanning an interval between $\sim$1 to 40 lt-days. These sizes can be converted to half-light radii by multiplying by 1.18, $R_{1/2}=1.18 r_s$.\\

\begin{figure}[h]
\centering
\includegraphics[width=8.5cm]{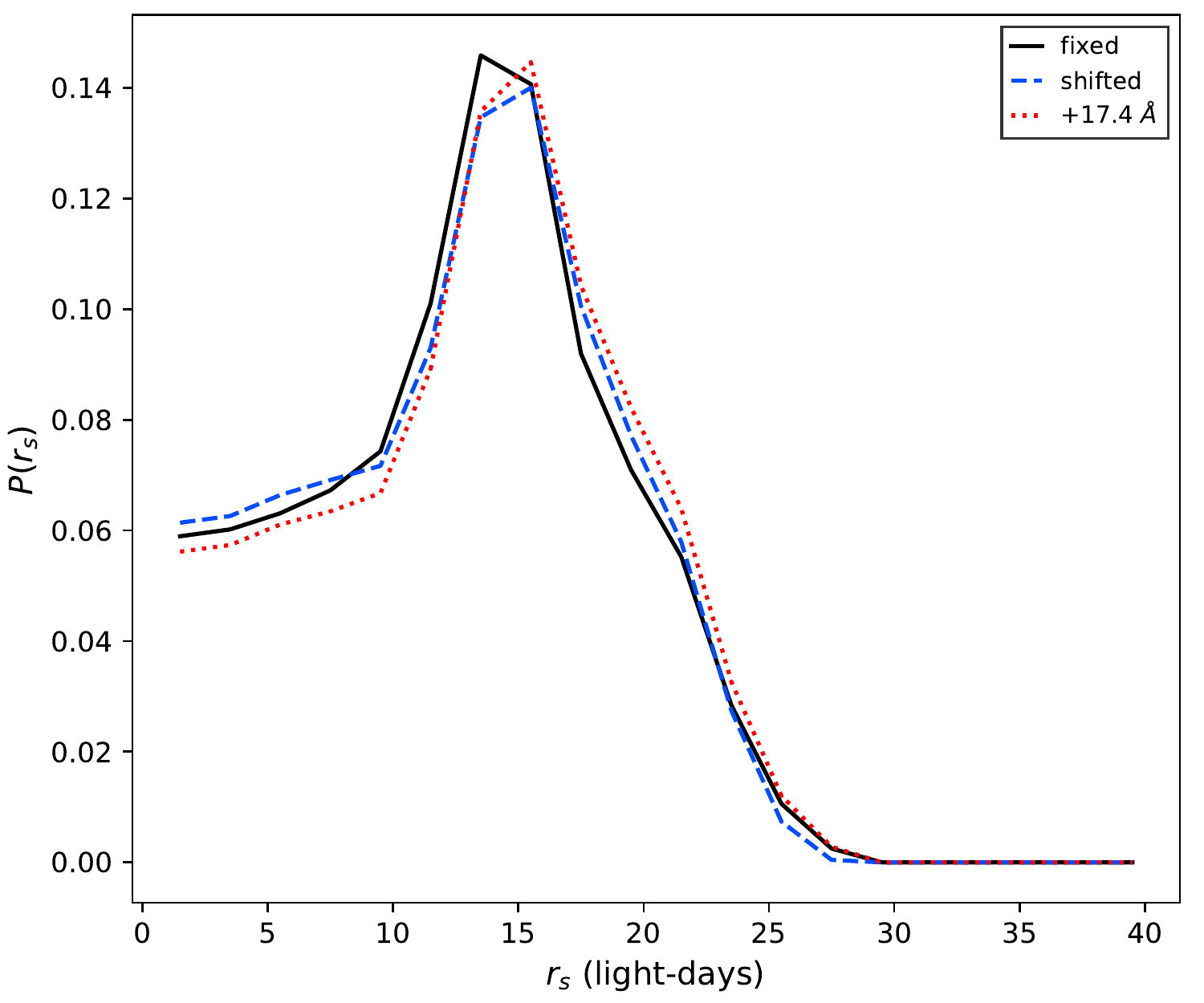}
\caption{Probability distribution of the region emitting the Fe III $\lambda\lambda$2039-2113 blend using different wavelength windows for estimating the magnitude differences $\Delta m$.}
\label{PDF}
\end{figure}

Given the estimates of the differential microlensing in the Fe III emission blend between images A and B, we can estimate the size of its emission region. We follow the steps described in \citet{Guerras2013,Fian2018,Fian2021} and treat each microlensing measurement as single epoch event. From the microlensing corresponding to all available epochs of observation, we compute the joint microlensing probability, $P(r_s)$, to obtain an average estimate of the source size. The resulting joint likelihood functions for the Fe III emission blend using three different wavelength windows can be seen in Figure \ref{PDF}. It is clearly visible that a moderate change of the definition of the wavelength interval does not introduce significant changes in the size estimates. From Figure \ref{PDF}, we can infer a size of $R_{1/2} \sim15$ lt-days (68\% confidence) for the region emitting the Fe III UV blend, indicating that this feature is formed close to or within the accretion disk, which (according to paper I and \citealt{Cornachione2020}) has a size of $\sim17.6$ lt-days at $\lambda_{rest} = 2558\AA$.

Under the hypothesis that the redshift of the FeIII $\lambda\lambda$2039-2113 is of gravitational origin, we can use the previously obtained redshift of the Fe III lines together with the inferred microlensing based size of the Fe III emitting region to derive the mass of the central SMBH. Table \ref{rm} lists the SMBH masses obtained after inserting the mean redshift, $\langle \Delta \lambda \rangle$, and the half-light radius of the Fe III blend (using different wavelength windows), $R_{1/2}=R_{FeIII}$, in Eq. \ref{mass}. The errors in the Fe III emitting region sizes are estimated by a maximum likelihood method. The uncertainties in the SMBH masses are proportional to the square root of the sum of the squared relative errors of the emitting region sizes and the squared standard error in the mean redshift, $\sigma_{M_{BH}}=\sqrt{\sigma_{R_{FeIII}}^2+\sigma_{\Delta \lambda}^2}$.
\begin{table}[h]
\renewcommand{\arraystretch}{1.4}
\caption{Half-light radii ($R_{1/2}$) inferred from Figure \ref{PDF} and corresponding SMBH mass ($M_{BH}$).}
\begin{tabu}to 0.49\textwidth {X[l]X[l]X[l]}

\hline
\hline
Interval & $R_{1/2}$ (lt-days) & $M_{BH}$ ($\times10^9 M_\odot$)\\
\hline 
Fixed & $15.0_{-3.5}^{+4.0}$ & $1.47_{-0.48}^{+0.52}$ \\
Shifted & $15.0_{-3.9}^{+4.1}$ & $1.47_{-0.50}^{+0.53}$ \\
+17.4$\AA$ & $15.5_{-3.6}^{+4.2}$ & $1.52_{-0.49}^{+0.53}$ \\
\hline 
\end{tabu}
\label{rm}
\end{table}

\subsection{Impact of stellar mass fraction on size estimates}
Since the amplitude of microlensing (and as a consequence the source size) is sensitive to the local stellar surface mass density function, we estimate the impact of the stellar mass fraction $\alpha$ on the Fe III emitting region size. We adopted a value $\alpha=0.3$ for image B and $\alpha=0$ for image A\footnote{The two lensed images of Q 0957+561 are located at very different radii from the center of the lens, resulting in different fractions of convergence in the form of stars.} (for more details see paper I), and repeated all the calculations. As expected, we obtain a somewhat larger size for the emitting region of Fe III, leading to a $\sim1.3$ times bigger SMBH mass of $M_{BH} \sim 2\times10^9 M_\odot$ (see Table \ref{alpha}).

\begin{table}[h]
\renewcommand{\arraystretch}{1.4}
\caption{Half-light radii ($R_{1/2}$) and corresponding SMBH mass ($M_{BH}$) for stellar mass fractions of $\alpha=0$ (image A) and $\alpha=0.3$ (image B).}
\begin{tabu}to 0.49\textwidth {X[l]X[l]X[l]}

\hline
\hline
Interval & $R_{1/2}$ (lt-days) & $M_{BH}$ ($\times10^9 M_\odot$)\\
\hline 
Fixed & $20.1_{-6.0}^{+5.5}$ & $1.94_{-0.68}^{+0.64}$ \\
Shifted & $20.1_{-6.5}^{+5.7}$ & $1.94_{-0.72}^{+0.65}$ \\
+17.4$\AA$ & $20.6_{-6.5}^{+2.5}$ & $2.01_{-0.72}^{+0.42}$ \\
\hline 
\end{tabu}
\label{alpha}
\end{table}

\subsection{Impact of spectroscopic sample on redshift estimates}
\begin{figure*}[h]
\centering
\includegraphics[width=13.7cm]{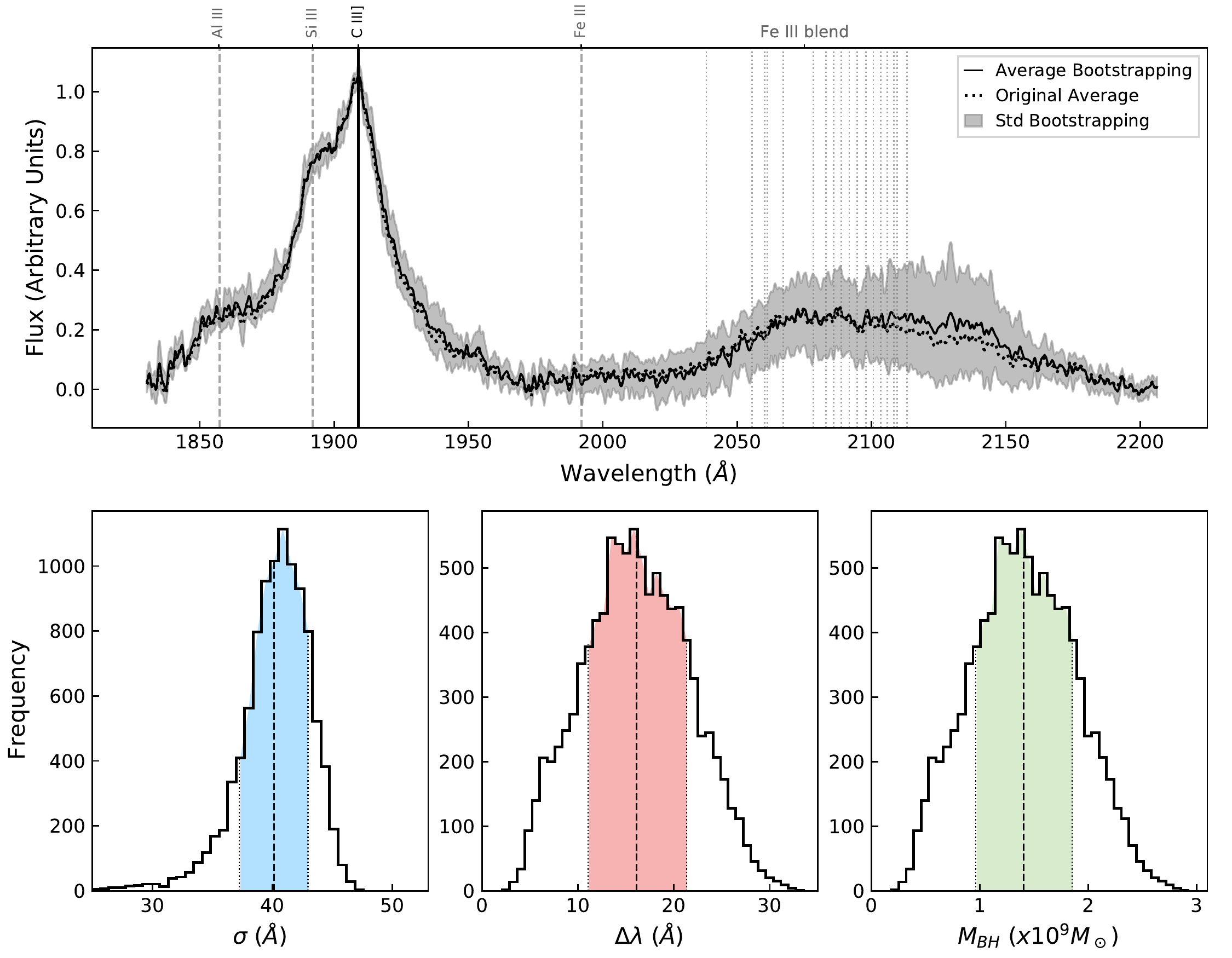}
\caption{Bootstrapping results. Top: Average spectrum (black solid line) and standard deviation (gray error envelope) obtained by using bootstrapping, together with the original average (black dotted line). Bottom: Frequency distributions of the broadening, $\sigma$, redshift, $\Delta \lambda$, and resulting SMBH mass, $M_{BH}$. The dashed vertical lines give the mean of each distribution and the colored shaded areas show $\pm1$ standard deviation.}
\label{bootstrap}
\end{figure*}

To evaluate the impact of single spectra on the average signal (and hence on the average redshift), we used bootstrapping to randomly pick spectra from our sample and build the average spectrum, fit the Fe III $\lambda\lambda$2039-2113 blend, and estimate the corresponding redshift. The simulated average spectrum is made
of different combinations (with repetition) of ten out of the original ten spectra. We repeated this process 10000 times to obtain frequency distributions for the broadening, $\sigma$, and the redshift, $\Delta \lambda$, which are shown in the lower panels of Figure \ref{bootstrap}. We find that we are biased towards a slightly bigger redshift ($\sim$1\AA\ on average) and thereby a somewhat bigger (but consistent within uncertainties) SMBH mass ($\Delta M_{BH} \sim 0.1 \times 10^9 M_\odot$, see Table \ref{bs}).

\begin{table}[h]
\renewcommand{\arraystretch}{1.4}
\caption{Values obtained from the bootstrap-based frequency distributions shown in Figure \ref{bootstrap}.}
\begin{tabu}to 0.49\textwidth {X[l]X[c]X[c]}

\hline
\hline
Parameter & Bootstrapping & Difference\\
\hline 
$\sigma$ (\AA) & $40.2 \pm 3.3$ & $1.2$ \\
$\Delta \lambda$ (\AA) & $16.1\pm5.5$ & $1.3$ \\
$M_{BH}\ (\times10^9 M_\odot$) & $1.40\pm0.48 $ & $0.12$ \\
\hline 
\end{tabu}
\label{bs}
\end{table}

\section{Discussion}\label{4}
Using the individual determination of $R_{FeIII}$ for Q 0957+561 and the mean redshift $\langle \Delta \lambda \rangle$ obtained by averaging the five epochs times the two images available for the spectra, we inferred a SMBH mass, $M_{BH}=1.5^{+0.5}_{-0.5} \times 10^9 M_\odot$. This value is in agreement within uncertainties with the virial determinations from \citet{Assef2011}, $M_{BH}=1.0^{+1.1}_{-0.5} \times 10^9 M_\odot$ (CIV) and with our previous measurement, $M_{BH}= (0.9\pm0.5) \times10^9 M_\odot$  (\citealt{Mediavilla2018}), obtained from a spectrum of relatively poor S/N ratio and based in an average estimate for $R_{FeIII}$. However, our new determination is affected by relatively small errors $\lesssim 35\%$, which represent a significant improvement in the context of SMBH mass measurements.\\

The strong variability of the Fe III $\lambda\lambda$2039-2113 feature observed in the spectral monitoring (see Figure \ref{epochs}) is evidence of the small size of the region emitting this blend. We observe a global variability of both images with respect to CIII] with maxima in 2008 and 2016, and an enhancement of component B in 2008 and 2009. Given the large lag ($417\pm2$ days; see, e.g., \citealt{Shalyapin2008}) between both images, in principle, it may be assumed that the difference between A and B at a given epoch could arise from a lag in the intrinsic variability of the quasar. However, the similarity of the spectra during 2008 and 2009 and the subsequent global fading of both components in 2010 and 2011 contradict this hypothesis. Thus, the microlensing hypothesis is favored by the observations to explain the differences between A and B while the quasar intrinsic variability may be the cause of the global changes. Based on the hypothesis of gravitational microlensing, we obtained a size of about 15 lt-days (equivalent to $\sim88$ Schwarzschild radii for a $1.5\times10^9 M_\odot$ black hole), which match  the size of the accretion disk inferred from the microlensing of the continuum very well (see paper I\color{black}). This agreement gives consistence to our results and support the origin of the Fe III $\lambda\lambda$2039-2113 feature in the accretion disk.\\

Finally, from the broadening of the H$\beta$ emission line (\citealt{Assef2011}), we can estimate the virial factor using the equation (\citealt{Mediavilla2019}):
\begin{equation}
\label{virialfactor}
 f_{H\beta} \simeq{2 \over 3}{R_{FeIII}\over R_{H\beta}}{\left({\Delta \lambda\over \lambda}\right)_{FeIII}\over \left(FWHM_{H\beta}/c\right)^2}.
\end{equation}
To obtain $R_{H\beta}$ we take the R-L scaling adopted by \citet{MejiaRestrepo2016}, 
\begin{equation}
\label{RHb}
R_ {H\beta}=538 \left({\lambda L_{\lambda5100}\over 10^{46} {\, \rm erg\,s^{-1}}}\right)^{0.65}\rm lt-days.
\end{equation}

Inserting the corresponding values (and using the continuum luminosity estimate of \citealt{Assef2011}, $\lambda L_{5100\AA} = 10^{45.79}$ erg s$^{-1}$) in Eq. \ref{virialfactor}, we finally obtain $f_{H\beta} =1.8$. This value is in good agreement with the empirical calibration recently proposed by \citealt{Yu2019} for the FWHM-based virial factor of $H\beta$ ($f_{H\beta} = 1.7$). For C IV, assuming $R_{C IV}\sim R_ {H\beta}$ and taking different prescriptions to define the line continuum (\citealt{Assef2011}), we derive $f_{CIV} =1.4 -1.8$. These values are comparable to the prediction, $f=1.2$, by \citealt{Collin2006} (see Eq. 6 of these authors) corresponding to the width estimate by \citet{Assef2011}, $FWHM_{H\beta}=3300\rm\, km\, s^{-1}$. For Fe III, we obtain a slightly bigger virial factor of $f_{FeIII}=2.3$ when using the FWHM of the Fe III blend inferred from our average spectrum. According to the simple model proposed by \citet{Collin2006}, this range of values corresponds to ratios between isotropic (turbulent) and cylindrical (Keplerian) velocities, $a={V_{iso}/V_{cyl}}\lesssim 0.4$, and inclinations, $i\lesssim 30^{\rm o}$.

\section{Conclusions}\label{5}

We used microlensing measurements of the Fe III $\lambda\lambda$2039-2113 blend to estimate a size of $R_{FeIII}\sim 15$ lt-days of the region emitting this feature in Q 0957+561. This result is in good agreement with the accretion disk dimensions inferred from microlensing of the continuum ($17.6_{-5.1}^{+3.4}$ lt-days according to paper I). Using this size and the redshift of the Fe III $\lambda\lambda$2039-2113 emission lines, we obtained a mass estimate, $M_{BH}=1.5^{+0.5}_{-0.5} \times 10^9 M_\odot$, which is in agreement to within errors with previous determinations. The relatively small uncertainties in the mass determination ($\lesssim 35\%$) make this method (Fe III $\lambda\lambda$2039-2113 redshift plus microlensing based size) a compelling alternative to other methods for the measurement of SMBH masses such as the virial plus RM-based size. Combining the Fe III $\lambda\lambda$2039-2113 redshift based method with the virial, for Q 0957+561 we estimate  a virial factor in the range of $f\sim 1.2-1.7,$  which may correspond to a BLR (broad-line region) with moderate to low isotropy ($a={V_{iso}/V_{cyl}}\lesssim 0.4$) and low inclination ($i\lesssim 30^{\rm o}$).\\

\begin{acknowledgements}
We thank the anonymous referee for the helpful comments and the constructive remarks on this manuscript. We thank the GLENDAMA project for making publicly available the spectroscopic data of Q 0957+561. C.F. gratefully acknowledges the financial support of the Tel Aviv University and University of Haifa through a DFG grant HA3555-14/1. E.M. and J.A.M are supported by the Spanish MINECO with the grants AYA2016- 79104-C3-1-P and AYA2016-79104-C3-3-P. J.A.M. is also supported from the Generalitat Valenciana project of excellence Prometeo/2020/085. J.J.V. is supported by the project AYA2014-53506-P financed by the Spanish Ministerio de Econom\'\i a y Competividad and by the Fondo Europeo de Desarrollo Regional (FEDER), and by project FQM-108 financed by Junta de Andaluc\'\i a. V.M. acknowledges partial support from Centro de Astrof\'\i sica de Valpara\'\i so (CAV). 
\end{acknowledgements}

\bibliographystyle{aa}
\bibliography{bib_paper}

\end{document}